\documentclass[epj]{svjour}

\usepackage{graphicx}
\usepackage{fancyhdr}
\def\makeheadbox{{
 }}
\def\mytitle{My title} 
\def\myauthors{My name}  
\def\mytype{My type of session}
\def\mysession{My session}

\def\mytitle{A light non-standard Higgs boson: to be or not to be 
at a (Super) B factory?} 
\def\myauthors{Miguel-Angel Sanchis-Lozano}    
\def\mytype{Contributed Talk}    
\def\mysession{Colliders - Higgs Phenomenology}

\begin{document}
\title{A light non-standard Higgs boson: to be or not to be 
at a (Super) B factory?}
\author{Miguel-Angel Sanchis-Lozano\inst{1}
\thanks{\emph{Email:} Miguel.Angel.Sanchis@uv.es}%
}                     
%
%
\institute{Departament de F\'{\i}sica Te\`orica and IFIC, 
Universitat de Val\`encia-CSIC, 46100 Burjassot, Spain}
%
\date{}
\abstract{A light non-standard Higgs boson decaying mainly into $\tau^+\tau^-$
has not been yet ruled out by LEP searches in several scenarios. 
We verify that, in the context of the Next-to-Minimal-Supersymmetric 
Standard Model, a low-mass CP-odd (mostly but not completely singlet-like) 
Higgs boson
can couple strongly enough to down-type fermions to be detected
in radiative $\Upsilon$ decays into tauonic pairs
at a high-luminosity B factory. Possible
spectroscopic effects of a mixing with $\eta_b$ resonances are
also analyzed.
} 
\maketitle
\section{Introduction}
\label{intro}
In spite of intensive searches performed at LEP, the possibility of 
a light non-standard Higgs boson has not been excluded yet in
several scenarios beyond the Standard Model (SM). Moreover, the
LHC might not be able to find a signal from a light Higgs boson whose mass is
below the $B\bar{B}$ threshold. A Super B factory can thus play an
important and complementary role in this regard \cite{Bona:2007qt}. 

From a theoretical viewpoint, the existence of a light
pseudoscalar Higgs is not unexpected in certain non-minimal
extensions of the SM. As an especially appealing example, the 
next-to-minimal supersymmetric standard 
model (NMSSM)  gets a gauge singlet added to the
MSSM two-doublet Higgs sector, leading
to seven physical Higgs bosons, five of them neutral including
two pseudoscalars \cite{gunion}.  
Interestingly, the authors of \cite{Dermisek:2005gg}
interpret, within the NMSSM, the excess of $Z$+$b$-jets events found at LEP  
as a signal of a 
SM-like Higgs decaying partly into $b\bar{b}\tau^+\tau^-$, but
dominantly into $\tau$'s via two light pseudoscalars.
Let us also mention the exciting connection with
possible light neutralino dark matter  
\cite{Gunion:2005rw} and its detection
at B factories \cite{McElrath:2005bp}.

The possibility of light Higgs particles can be extended to 
scenarios with more than one gauge singlet \cite{Han:2004yd}, 
and even to the MSSM with a CP-violating Higgs sector 
\cite{Carena:2002bb,Lee:2007ai} as 
LEP bounds can be evaded \cite{Kraml:2006ga,Dermisek:2006py}.
In the CP-violating benchmark scenario and several variants,  
the combined LEP data show large domains of the parameter space
which are not excluded, down to the lowest Higgs mass values  
\cite{Schael:2006cr}. A similar conclusion applies to  
a two Higgs doublet model of type II (2HDM(II)) \cite{gunion}, where 
some windows for a very light Higgs are still open
\cite{Abbiendi:2004gn}. In addition, Little Higgs models have 
an extended structure 
of global symmetries
(among which there can appear $U(1)$ factors) broken
both spontaneously and explicitly, leading to possible light pseudoscalar
particles in the Higgs spectrum \cite{Kraml:2006ga}.
Finally, let us mention the $g-2$ muon anomaly  
that might require
a light CP-odd Higgs boson \cite{Krawczyk:2002df}
to reconcile the experimental value with the SM result
\cite{Hagiwara:2006jt}. 

\section{A light CP-odd Higgs in the NMSSM?}
\label{sec:2}

As is well-known, the NMSSM provides an elegant solution to the $\mu$ problem
of the MSSM via the introduction of a singlet superfield $\hat{S}$
in the Higgs sector. 
As compared to the three independent parameters of the MSSM (usually
chosen as $\tan{\beta}$, $\mu$ and $M_A$), the Higgs sector of
the NMSSM requires six parameters, namely, $\lambda,\ \kappa,\ A_{\lambda},
\ A_{\kappa},\ \tan{\beta},\mu_{eff}$, where $\mu_{eff}=\lambda s$ 
is the effective $\mu$-term generated from the
vev of the singlet field, $s \equiv <S>$; $\lambda A_{\lambda}$ 
and $\kappa A_{\kappa}$ appear in the trilinear
soft-supersymmetric-breaking terms of the potential. Our sign
conventions are: $\lambda$ and $\tan{\beta}$ always positive,
while $\kappa,\ A_{\lambda},\ A_{\kappa},\ \mu_{eff}$
are allowed to have either sign.
In the limit of either
slightly broken $R$ or Peccei-Quinn (PQ) symmetries,
the lightest (CP-odd) Higgs boson 
\footnote{The lightest CP-odd Higgs will be denoted as $A^0$ throughout 
this work instead of the more common $A_1$ in the NMSSM pointing out 
that a light pseudocalar Higgs-like particle might also exist
in other scenarios.}
can be much lighter than the other Higgs bosons.

The non-singlet fraction of the $A^0$ is defined by $\cos{\theta_A}$: 
\[ A^0=\cos{\theta_A}A_{MSSM}+\sin{\theta_A}A_s \] 
where $\theta_A$ is the mixing angle between the singlet
component and the MSSM-like component of the $A^0$.

The $2 \times 2$ square mass matrix for the CP-odd Higgs bosons
has the following matrix elements \cite{Ellwanger:2004xm}
\begin{eqnarray}
M_{11}^2 &=& \frac{2 \lambda s}{\sin{2\beta}}(A_{\lambda}+\kappa s),\ 
M_{12}^2\ =\  \lambda v (A_{\lambda}-2 \kappa s) \nonumber \\ 
M_{22}^2 &=&2 \lambda \kappa v^2 \sin{2\beta}+\lambda A_{\lambda}\frac{v^2\sin{2\beta}}{2s}-3 \kappa A_{\kappa}s \nonumber
\end{eqnarray}
Defining $\Delta M^2=\sqrt{(M_{11}^2-M_{22}^2)^2+4(M_{12}^2)^2}$, the
eigenstate mass of the lightest CP-odd Higgs boson can be written as 
\begin{equation} 
m_{A^0}^2=\frac{1}{2}[M_{11}^2+M_{22}^2 - \Delta M^2] 
\end{equation}
and the mixing angle reads 
\begin{equation}
\cos{\theta_A}=-\frac{M_{11}^2-M_{22}^2}{\Delta M^2}
\end{equation}

For small trilinear couplings, 
the mass of the lightest CP-odd Higgs boson 
and $\cos{\theta_A}$ can be approximated by
\begin{eqnarray}
m_{A^0}^2 &\simeq& 3s\ \biggl(\frac{3 \lambda A_{\lambda}}{2 \sin{2 \beta}}
\cos{}^2\theta_A-\kappa A_{\kappa}\sin{}^2\theta_A\biggr) \\
\cos{\theta_A} &\simeq& -\frac{\lambda v(A_{\lambda}-2 \kappa s)\sin{2\beta}}
{2\lambda s (A_{\lambda}+\kappa s)+3 \kappa A_{\kappa}s \sin{2\beta}} 
\end{eqnarray}

From Eqs.(3-4) one can see that, under the protective symmetries 
mentioned in the Introduction, a small
$A^0$ mass can be achieved in two ways:
\begin{itemize} 
\item[$(i):$] $|\cos{\theta_A}| \simeq 0$ and $\sin{\theta_A} \simeq 1$, 
the $A^0$ is almost entirely singlet and the CP-odd Higgs mass
is given from Eq.(3) by 
$m_{A^0}^2 \simeq -3 \kappa A_{\kappa}s$, 
which can be very small under a PQ symmetry. Likewise the $A^0$ coupling
to down-type fermions  ($\sim \cos{\theta_A}$ $\tan{\beta}$)
would remain small even at large $\tan{\beta}$ since 
$\cos{\theta_A} \sim \sin{2\beta} \simeq 2/\tan{\beta}$.

\item[$(ii):$] $|\cos{\theta_A}|$ is not so small 
(e.g. $|\cos{\theta_A}| \simeq 0.1-0.5$)
but still keeping the mostly singlet nature of the $A^0$. 
The two terms inside the bracket  of
Eq.(3) tend to cancel (especially
for large $\tan{\beta}$ due to the $\sin{2\beta}$ in 
the denominator of the first term) provided that $\lambda A_{\lambda}$ and 
$\kappa A_{\kappa}$ have the same sign, resulting in a low
$m_{A^0}$ value \cite{Dermisek:2007yt}. This possibility should be realized 
along the straight line on the $\lambda-\kappa$ plane
where $|\cos{\theta_A}|$ is enhanced but
keeping, we insist, the singlet character of the $A^0$ to a
large extent (at the $\sim 1\%$ probability level).
\end{itemize}

Therefore, following $(ii)$,
a low $A^0$ mass (e.g. $m_{A^0} < 2m_b$) can easily emerge 
at large $\tan{\beta}$ with, at the same time, a
fairly enhanced coupling to $b$ quarks and $\tau$ leptons,
yielding observable effects in $\Upsilon$ decays
as advocated in a series of papers 
\cite{Sanchis-Lozano:2006gx,Sanchis-Lozano:2005di,Sanchis-Lozano:2003ha}.
In fact, different sets of NMSSM parameters can lead to 
this possibility, notably those yielding moderate 
$|\lambda A_{\lambda}|$ and small $|\kappa A_{\kappa}|$ values
(evaluated at the scale $m_Z$) with $\mu_{eff} \simeq 150$ GeV,
likely corresponding to 
the smallest degree of fine-tuning according to the analysis of LEP 
Higgs event excess \cite{Dermisek:2005ar,Dermisek:2007yt}.

\begin{figure}
\begin{center}
\includegraphics[width=14pc]{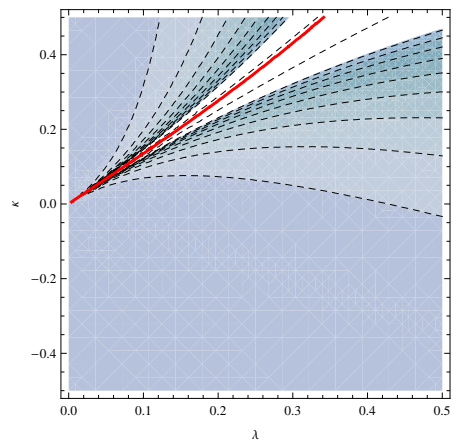}
\caption{Contour dashed lines for $X_d=0.5,1,...,4.5$
on the $\lambda-\kappa$ plane from Eq.(2) setting 
$A_{\lambda}=-200$, 
$A_{\kappa}=-15$, $\mu_{eff}=150$ (at the scale $m_Z$, all in GeV)
and $\tan{\beta}=50$. The unshaded area stands
for $X_d \geq 5$; the innermost dashed 
line corresponds to $X_d=10$ while the thick red line represents 
$m_{A^0}=10$ GeV.}
\end{center}
\label{fig:1}       
\end{figure}

Defining $X_d=\cos{\theta_A}\tan{\beta}$, 
the contour lines on the $\lambda-\kappa$ plane
for $X_d=0.5,1,...,4.5$ and $X_d=10$ are displayed in Fig.1 
(the unshaded area standing for $X_d \geq 5$, 
when $\Upsilon$ leptonic decays should start to become sensitive to the 
$A^0$-mediated annihilation channels). In our
plot we employed Eq.(2)    
setting as reference values: $A_{\lambda}=-200$ GeV, 
$A_{\kappa}=-15$ GeV and $\mu_{eff}=150$ GeV,
with $\tan{\beta}=50$. Let us remark that
somewhat smaller values of $\tan{\beta}$ ($\tan{\beta} > 25$) lead to 
qualitatively similar plots.

Indeed one could expect from Eq.(4)
an enhancement of $|\cos{\theta_A}|$ (and therefore of $|X_d|$)
in the vicinity of 
the straight line defined by $\lambda A_{\lambda}+\kappa\ \mu_{eff}=0$ since
then both $\sin{2\beta}$ terms tend to cancel 
in the ratio (exactly cancelling out along the straight line). Thus,
the product $|\cos{\theta}_A\tan{\beta}|$ can further increase
at large $\tan{\beta}$, although limited by 
experimental bounds if the coupling to $b$ quarks becomes too large.
Values of $|X_d| \sim 10$ are still
allowed in the NMSSM \cite{Hiller:2004ii}.

The thick red line in Fig.1 stands for $m_{A^0}=10$ GeV obtained
from Eq.(1), lying close to the $X_d=10$ contour-line and 
expectedly leading to a slight but observable lepton universality (LU) 
breakdown in $\Upsilon$ decays, to be commented in Sect.~\ref{sec:4}.

\section{Mixing of the $A^0$ and $\eta_b$ resonances}
\label{sec:3}

The mixing between a CP-odd Higgs and a $\eta_b$ resonance is
described by the introduction of off-diagonal
elements denoted by $\delta m^2$ in the mass matrix
\cite{Drees:1989du} 
\[
{\cal M}_0^2=  
\left(
     \begin{array}{cc}
      m_{A_0^0}^2-im_{A_0^0}\Gamma_{A_0^0} & \delta m^2\\
      \delta m^2 & m_{\eta_{b0}}^2-im_{\eta_{b0}}\Gamma_{\eta_{b0}}      
     \end{array}
\right)
\] 
where the subindex $\lq$0' indicates unmixed states: $m_{A_0^0}$
($\Gamma_{A_0^0}$) and $m_{\eta_{b0}}$ ($\Gamma_{\eta_{b0}}$) denote the 
masses (widths) of the pseudocalar Higgs boson and resonance, 
respectively. 

The $A^0$ and $\eta_b$ physical (mixed) states can be written as
\begin{eqnarray}
A^0 &=& \cos{\alpha}\ A_0^0\ +\ \sin{\alpha}\ \eta_{b0}
\nonumber \\
\eta_b &=& \cos{\alpha}\ \eta_{b0}\ -\ \sin{\alpha}\ A_0^0 
\nonumber
\end{eqnarray}
assuming $[|\cos{\alpha}|^2+|\sin{\alpha}|^2]^{1/2}\simeq 1$.
The definition of the mixing angle $\alpha$ and a lengthier discussion
can be found in \cite{Fullana:2007uq} (and references therein).

The off-diagonal element $\delta m^2$ can be computed (see Fig.1c)
within the framework of a nonrelativistic quark potential model to be
$\delta m^2$(GeV$^2$) $\approx\ 0.146\ \times\ X_d$ \cite{Fullana:2007uq}.
Notice that $\delta m^2$ is proportional to $X_d$.

\begin{figure}
\begin{center}
\includegraphics[width=12pc]{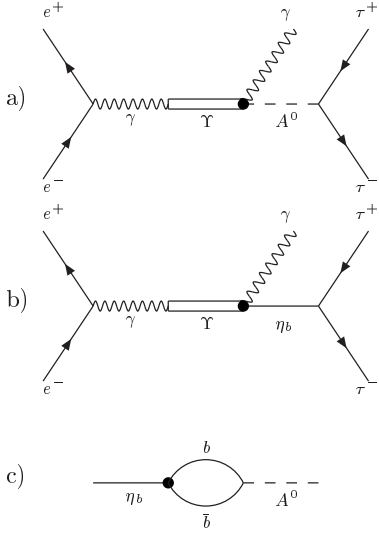}
\end{center}
\caption{Process $e^+e^-\to \Upsilon \to \gamma\ \tau^+\tau^-$ 
with a) pseudoscalar Higgs, b) $\eta_b$, as
intermediate states; c) Mixing diagram}
\end{figure}

\subsection{Radiative decay of the $\Upsilon$ into $\tau^+\tau^-$}

In Ref.\cite{Fullana:2007uq} we employed the mixing formalim to describe both
resonant and non-resonant $\Upsilon$ decays into
a photon and a tauonic pair as depicted in Fig.2.

The couplings 
of the physical $A^0$ and $\eta_b$ states to a $\tau^+\tau^-$ pair 
are given by
\begin{eqnarray}
g_{A^0\tau\tau} &=& 
\cos{\alpha}\ g_{A_0^0\tau\tau}+ 
\sin{\alpha}\ g_{\eta_{b0}\tau\tau}
\nonumber \\
g_{\eta_b\tau\tau} &=& \cos{\alpha}\ g_{\eta_{b0}\tau\tau}- 
\sin{\alpha}\ g_{A_0^0\tau\tau} 
\nonumber
\end{eqnarray}
The full widths $\Gamma_{A^0}$ and
$\Gamma_{\eta_b}$ of the physical states
can also be expressed in terms of the widths of the unmixed states
according to the simple formulae:
\begin{eqnarray}
\Gamma_{A^0} &=& |\cos{\alpha}|^2\ \Gamma_{A_0^0}\ +\  
|\sin{\alpha}|^2\ \Gamma_{\eta_{b0}}
\\
\Gamma_{\eta_b} &=& |\cos{\alpha}|^2\ \Gamma_{\eta_{b0}}\ +\  
|\sin{\alpha}|^2\ \Gamma_{A_0^0}
\end{eqnarray}

In the SM we can very approximately set $g_{\eta_{b0}\tau\tau} \simeq 0$ and 
thus 
$g_{A^0\tau\tau} \simeq g_{A_0^0\tau\tau}\cos{\alpha},\ \ g_{\eta_b\tau\tau}
\simeq -g_{A_0^0\tau\tau}\sin{\alpha}$
where $g_{A_0^0\tau\tau}$ can be obtained from the Yukawa coupling strength
\cite{Fullana:2007uq}. Therefore, in our description
of the process shown in Fig.2b, the $\eta_b$ actually
decays into $\tau^+\tau^-$ through its mixing with the $A^0$ boson.

\subsection{Spectroscopic effects}

In addition, spectroscopic effects can appear in 
$b\bar{b}(^1S_0)$ states of the bottomonium family if 
the $A^0_0-\eta_{b0}$ mixing sizeably shifts the
masses of the physical states. In Fig.3 we plot
$m_{\eta_b(1S)}-m_{\eta_{b0}(1S)}$ versus 
$m_{A_0^0}-m_{\eta_{b0}}$. Such a shift has to be
added (with its sign) to the QCD expected 
$m_{\Upsilon(1S)}-m_{\eta_b(1S)}$ hyperfine splitting,
whose theoretical prediction is achieving
a remarkable precision \cite{Brambilla:2004wf}. 
As a consequence, if $m_{\eta_{b0}}<m_{A_0^0}$,  
the  hyperfine splitting could increase considerably 
with respect to the SM expectations even at moderate $|X_d|$. 
On the contrary, if  $m_{\eta_{b0}}>m_{A_0^0}$ 
the observed hyperfine splitting should shrink,
and even the $\Upsilon(nS)$ and $\eta_b(nS)$ mass levels might be 
reversed at large enough $|X_d|$. However, this spectacular effect,
if overlooked, would paradoxically render hard the 
experimental observation of such a $\eta_b$ state!

\begin{figure}
\includegraphics[width=16pc]{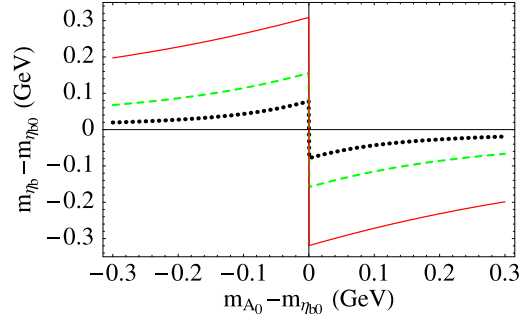}
\caption{Shift of the $\eta_b$ physical mass
induced by the mixing with a pseudoscalar Higgs boson, 
versus $m_{A_0^0}-m_{\eta_{b0}}$ for: 
dotted (black) line: $X_d=10$, dashed (green) line: $X_d=20$,
solid (red) line: $X_d=40$. The mass of the $\eta_b$ 
mixed state is decreased (increased) if $m_{A_0^0}>m_{\eta_{b0}}$
($m_{A_0^0}<m_{\eta_{b0}}$), ultimately implying a larger (smaller) 
$\Upsilon-\eta_b$ mass splitting than expected in the SM.}
\label{fig:2}       
\end{figure}

\section{Testing Lepton Universality in $\Upsilon$ decays}
\label{sec:4}

As emphasized in previous work (see \cite{Sanchis-Lozano:2006gx}
and references therein),  
the new physics contribution would be unwittingly ascribed to the 
$\Upsilon$ tauonic channel thereby breaking LU
if the (not necessarily soft) radiated photon 
escapes undetected in the experiment (the leptonic width is, in fact, 
an inclusive quantity with a sum over an infinite number of photons).

Experimentally, the relative importance of the Higgs-mediated channel
can be assessed via the ratio
\begin{equation}
{\cal R}_{\tau/\ell}= \frac{{\cal B}_{\tau\tau}-{\cal B}_{\ell\ell}}
{{\cal B}_{\ell\ell}}= \frac{{\cal B}_{\tau\tau}}{{\cal B}_{\ell\ell}}-1
\label{eq:R}
\end{equation}
where ${\cal B}_{\tau\tau}$ and ${\cal B}_{\ell\ell}$ denote the tauonic 
and ($\ell=e$) electronic or ($\ell=\mu$) muonic branching fractions of the 
$\Upsilon$ resonance, respectively. 
A (statistically significant) non-null value of
${\cal R}_{\tau/\ell}$ would imply the rejection
of LU (predicting ${\cal R}_{\tau/\ell} \simeq 0$)
and a strong argument suggesting
the existence of a pseudoscalar Higgs boson
mediating the process as shown in Fig.2. 

A thorough discussion of the physics underlying those diagrams,
useful expressions, values of the physical parameters and the
corresponding numerical analysis providing
$R_{\tau/\ell}$ as plotted in the set of figures 4, can be
found in Ref.\cite{Fullana:2007uq}.

\begin{figure}
\begin{center}
\includegraphics[width=16pc]{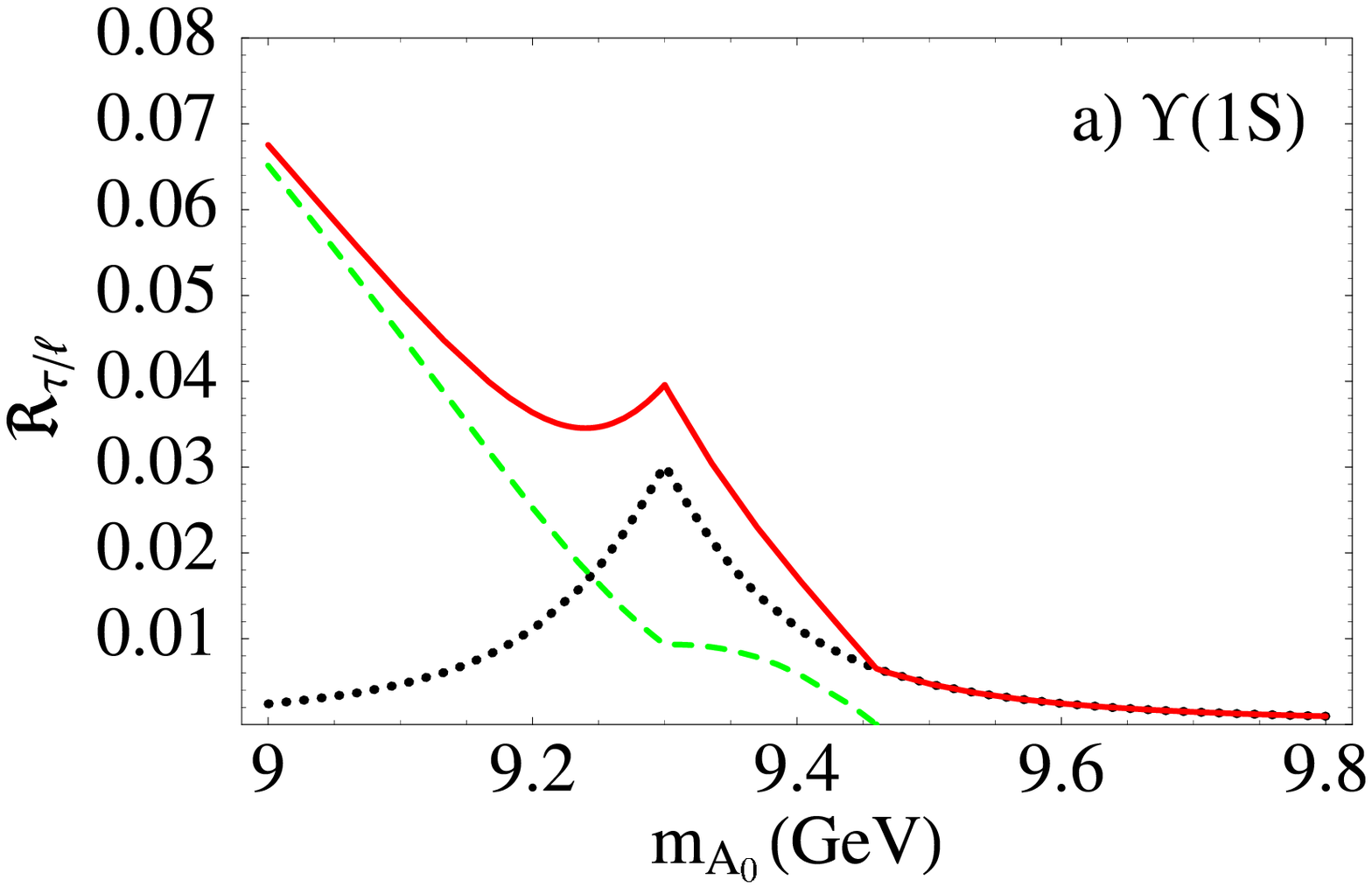}
\includegraphics[width=16pc]{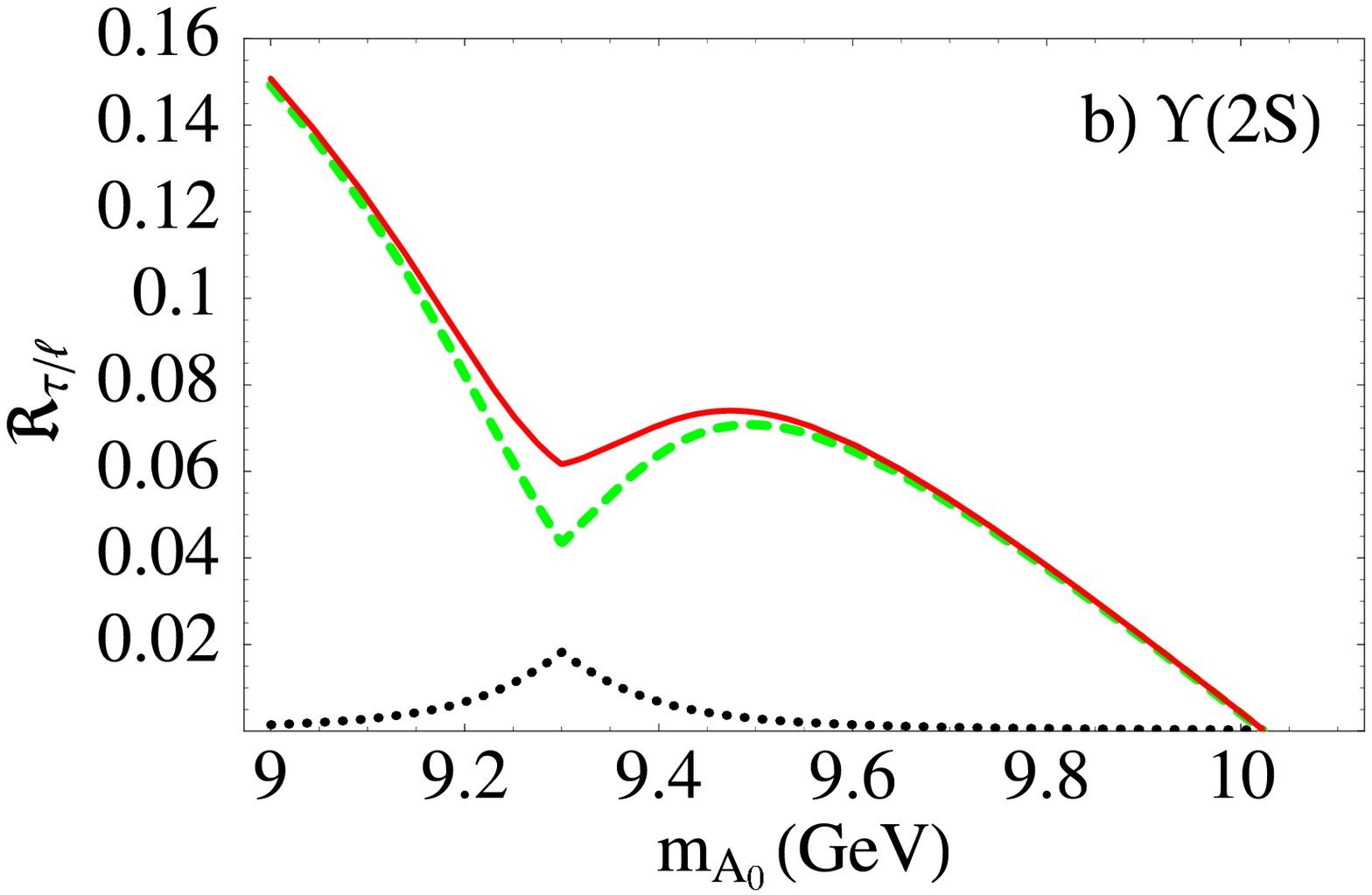}
\includegraphics[width=16pc]{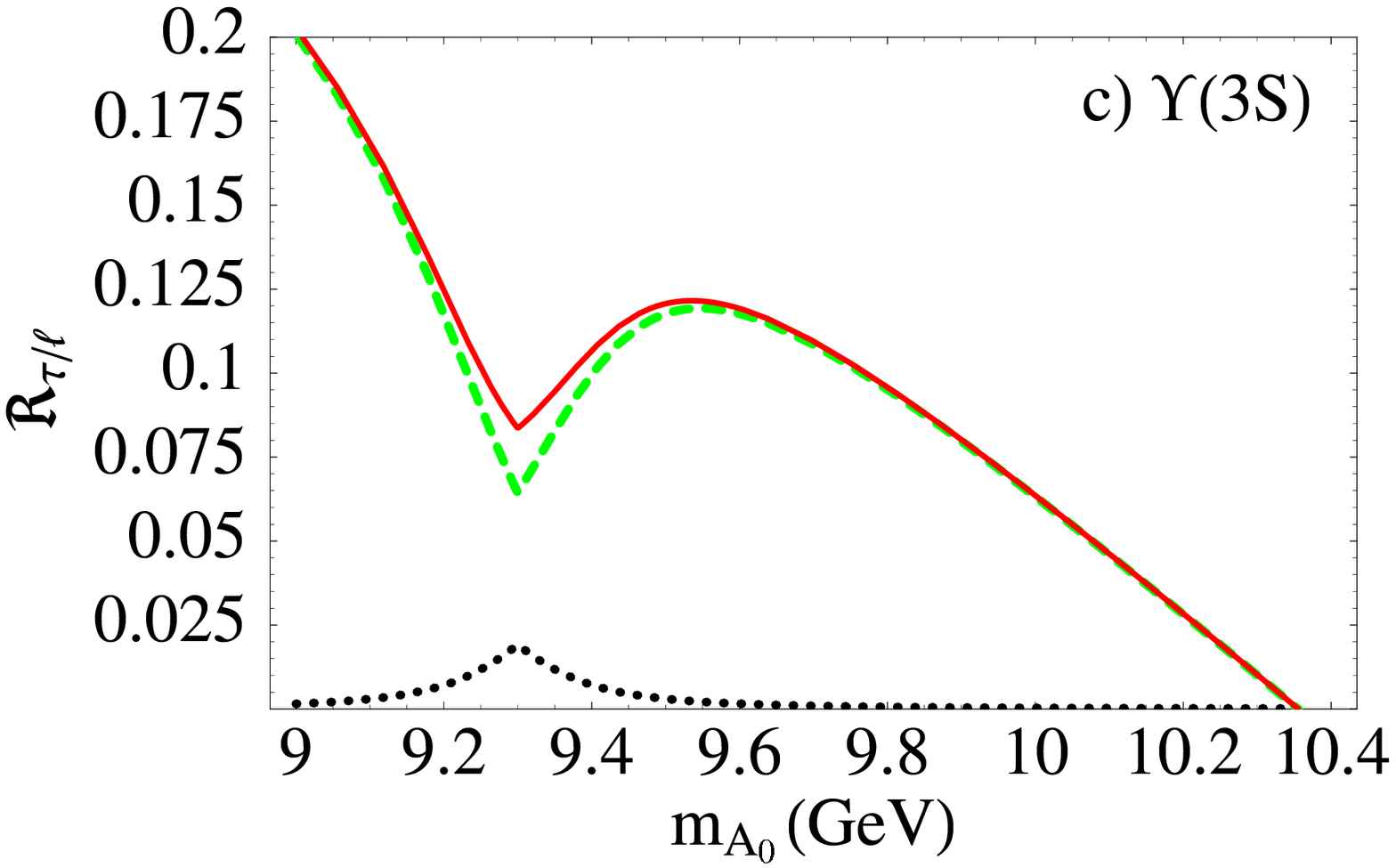}
\end{center}
\caption{$R_{\tau/\ell}$ versus the pseudoscalar Higgs mass
for $a)$ $\Upsilon(1S)$,  
$b)$ $\Upsilon(2S)$, and $c)$ $\Upsilon(3S)$ decays 
using $X_d=10$, $m_{\eta_{b0}}=9.3$ GeV, 
and $\Gamma_{\eta_{b0}}=5$ MeV, respectively. Resonant 
(dotted black line) and non-resonant (dashed green line)
decays are added in the solid red line. Larger (smaller) values 
of $X_d$ obviously yield higher
(lower) expectations for $R_{\tau/\ell}$.}
\label{fig:3}       
\end{figure}

By inspection, a bump can be observed in 
Fig.4a due to the
resonant contribution, while a dip appears in Figs.4b and 4c 
on account of the suppressed non-resonant channel, not
compensated by the resonant channel. 
In spite of that, the
higher $R_{\tau/\ell}$ values obtained 
for the $\Upsilon(2S)$ and $\Upsilon(3S)$ 
(due to the dominant
Wilczek mechanism of Fig.2a) allow us to conclude that
radiative decays of the latter resonances look more 
promising than
the $\Upsilon(1S)$ decays for the experimental observation 
of LU breaking at the few percent level at a B factory.
This conclusion is important if
a specific test of LU were to be put forward by
experimental collaborations \cite{Bona:2007qt}.

\section{Conclusions}

A mass of about 10 GeV can be easily obtained 
for the lightest CP-odd Higgs boson in the NMSSM with values
of $|X_d|=|\cos{\theta_A}| \times \tan{\beta} \simeq 10$, 
at large $\tan{\beta}$. The $A^0$ would still keep 
its predominantly singlet nature ($\cos{}^2\theta_A \sim$ few $\%$),
at the same time allowing a sizeable enhancement of its Yukawa coupling 
to down-type fermions (though escaping LEP bounds). Thus, we 
emphasize the relevance
of testing LU in $\Upsilon$ decays
to the few percent level
at presently running B factories, and the role to
be played by a future Super Flavor factory if this
effect were confirmed.\\

\noindent
{\em Acknowledgments}:
I acknowledge the perfect organization of the SUSY07
Workshop and the enjoyable atmosphere. I thank R. Dermisek and
J. Gunion for comments. Research under grants: 
FPA2005-01678 and \newline GVA COMP2007.



\begin{thebibliography}{999}
%
%



\bibitem{Bona:2007qt}
  M.~Bona {\it et al.},
  arXiv:0709.0451 [hep-ex].


\bibitem{gunion} J.~F.~Gunion, H.~E.~Haber, G.~Kane and S.~Dawson, 
\textit{The Higgs Hunter's Guide}
(Addison-Wesley Publishing Company, Redwood City, CA, 1990).





\bibitem{Dermisek:2005gg}
  R.~Dermisek and J.~F.~Gunion,
  Phys.\ Rev.\ D \textbf{73} (2006) 111701.



\bibitem{Gunion:2005rw}
  J.~F.~Gunion, D.~Hooper and B.~McElrath,
  Phys.\ Rev.\ D \textbf{73} (2006) 015011. 


\bibitem{McElrath:2005bp}
  B.~McElrath,
  Phys.\ Rev.\ D {\bf 72}, 103508 (2005).



\bibitem{Han:2004yd}
  T.~Han, P.~Langacker and B.~McElrath,
  Phys.\ Rev.\ D {\bf 70} (2004) 115006.



\bibitem{Carena:2002bb}
  M.~Carena, J.~R.~Ellis, S.~Mrenna, A.~Pilaftsis and C.~E.~M.~Wagner,
  Nucl.\ Phys.\ B \textbf{659} (2003) 145.


\bibitem{Lee:2007ai}
  J.~S.~Lee and S.~Scopel,
  Phys.\ Rev.\  D {\bf 75} (2007) 075001
  [arXiv:hep-ph/0701221].



\bibitem{Kraml:2006ga}
  S.~Kraml \textit{et al.},
  arXiv:hep-ph/0608079.



\bibitem{Dermisek:2006py}
  R.~Dermisek, J.~F.~Gunion and B.~McElrath,
  arXiv:hep-ph/0612031.



\bibitem{Schael:2006cr}
  S.~Schael {\it et al.}  [ALEPH Collaboration],
  Eur.\ Phys.\ J.\  C {\bf 47} (2006) 547.


\bibitem{Abbiendi:2004gn}
  G.~Abbiendi {\it et al.}  [OPAL Collaboration],
  Eur.\ Phys.\ J.\  C {\bf 40} (2005) 317.


\bibitem{Krawczyk:2002df}
  M.~Krawczyk,
  Acta Phys.\ Polon.\  B {\bf 33} (2002) 2621
  [arXiv:hep-ph/0208076].



\bibitem{Hagiwara:2006jt}
  K.~Hagiwara, A.~D.~Martin, D.~Nomura and T.~Teubner,
  arXiv:hep-ph/0611102.

\bibitem{Ellwanger:2004xm}
  U.~Ellwanger, J.~F.~Gunion and C.~Hugonie,
  JHEP {\bf 0502} (2005) 066
  [arXiv:hep-ph/0406215].


\bibitem{Dermisek:2007yt}
  R.~Dermisek and J.~F.~Gunion,
  arXiv:0705.4387 [hep-ph].


\bibitem{Sanchis-Lozano:2006gx}
  M.~A.~Sanchis-Lozano,
  J.\ Phys.\ Soc.\ Jap.\  {\bf 76} (2007) 044101
  [arXiv:hep-ph/0610046].


\bibitem{Sanchis-Lozano:2005di}
  M.~A.~Sanchis-Lozano,
  PoS \textbf{HEP2005} (2006) 334.


\bibitem{Sanchis-Lozano:2003ha}
  M.~A.~Sanchis-Lozano,
  Int.\ J.\ Mod.\ Phys.\ A \textbf{19} (2004) 2183.






\bibitem{Dermisek:2005ar}
  R.~Dermisek and J.~F.~Gunion,
  Phys.\ Rev.\ Lett.\  {\bf 95} (2005) 041801.



\bibitem{Hiller:2004ii}
  G.~Hiller,
  Phys.\ Rev.\ D {\bf 70}, 034018 (2004).


\bibitem{Drees:1989du}
M.~Drees and K.~i.~Hikasa,
Phys.\ Rev.\ D {\bf 41}, 1547 (1990).


\bibitem{Fullana:2007uq}
  E.~Fullana and M.~A.~Sanchis-Lozano,
  Phys.\ Lett.\  B {\bf 653} (2007) 67
  [arXiv:hep-ph/0702190].



\bibitem{Brambilla:2004wf}
  N.~Brambilla {\it et al.},
  arXiv:hep-ph/0412158.





\end{thebibliography}
\end{document}